\documentclass[aps,prb,amssymb,twocolumn]{revtex4}   

\newcommand{\be}{\begin{eqnarray}}
\newcommand{\ee}{\end{eqnarray}}

\begin{document}
\draft
\title{Magnetic disorder and dynamical properties of a Bose-Einstein
condensate in atomic waveguides}

\author{Daw-Wei Wang}

\address{
Department of Physics, National Tsing-Hua University, Hsinchu, Taiwan, ROC}
\date{\today}

\begin{abstract}
We systematically investigate the properties of 
the quenched disorder potential in an atomic waveguide, and study its
effects to the dynamics of condensate in the strong disorder region.
We show that even very small wire shape fluctuations can cause strong
disorder potential along the wire direction, leading to the fragmentation 
phenomena as the condensate is close to the wire surface. 
The generic disorder potential is Gaussian correlated random 
potential with vanishing
correlations in both short and long wavelength limits and with a strong
correlation weight at a finite length scale, set by the atom-wire distance.
When the condensate is fragmentized, we investigate the coherent and incoherent
dynamics of the condensate, and demonstrate that it
can undergo a crossover from a coherent condensate to an insulating Bose-glass
phase in strong disorder (or low density) regime. Our numerical results
obtained within the meanfield approximation 
are semi-quantitatively consistent with the experimental results.
\end{abstract}


\maketitle
\section{Introduction}

Low dimensional physics has long been an important and extensively studied
subjects in condensed matter physics since the early theoretical 
interest in 60th. In addition to the ordinary solid state system, e.g.
semi-conductor quantum wells, quantum wires, carbon nanotubes, and 
organic conductors etc., systems of ultracold atoms in highly anisotropic 
magneto-optical traps have been also a promising new system for the
low dimensional physics of both fermion and boson particles
\cite{ketterle,petrov,stoof}. Among varies proposals of the
anisotropic magneto-optical traps, microfabricated magnetic trap,
also called atomic waveguide or microtrap \cite{microchip_general}, 
has an additional important advantage for the coherent transport of atoms
along the quasi-one-dimensional (Q1D) waveguide potential
(for a typical system setup, see Fig. \ref{microchip}(a) and
the description in the next section).
Such combination of ultracold atoms in quantum mechanical limit with
the great versatility of the fabrication technique 
has opened a new direction to study many Q1D quantum physics, like
coherent transport \cite{aaron1}, beam splitter \cite{beamsplitter},
interference of matter wave \cite{interference}, and Tonks gas limit 
\cite{tonks}.
Many theoretical works have been proposed in the literature
to study these aspects in very recent years.

From experimental point of view, however, in order 
to successfully trap ultracold atoms in the atomic
waveguide, one has to apply very strong confinement
potential in the transverse dimension, which is usually achieved by
moving atoms very close to the conducting wire on the substrate.
(Electric current in the wire cannot be too large in order to avoid
heating \cite{private}.)
When the atom-wire distance is smaller than some critical values,
which is usually about $100-200$ $\mu$m,
however, nontrivial atom density modulation (fragmentation)
of the condensate occurs, as have been observed by different groups
\cite{aaron1,aaron2,zimmermann1,kraft02,hinds}.
It is also found that atom could cannot be transported
without excitations to higher
energy band when such fragmentation occurs inside the cloud 
\cite{aaron1,transport}.
For a static fragmentized atom cloud, 
it shows a rather generic (but not universal) length scale, $\lambda$, 
in $z$ direction, which becomes smaller when $d$ is reduced.
Kraft {\it et. al.} \cite{kraft02} further observe
that the positions of atom density maximum/minimum can be exchanged
if the direction of the offset magnetic field is reversed, showing
a nontrivial quenched disorder effects in the microtrap system.
Such interesting disorder effects can be very crucial when
considering the coherent transport in the magnetic waveguide,
and was first investigated theoretically by us recently \cite{early_work}.
We note that such quenched disorder induced 
fragmentation phenomena is different from the thermal fluctuation
studied in Ref. \cite{henkel}. The thermal 
fluctuation effects can be neglected in our present scope of interest,
because it becomes prominent only when the atoms are 
even closer to the wire surface (say $d<20$ $\mu$m), while
the fragmentation phenomena occur at well-larger distance 
(say $d<200$ $\mu$m).

The disorder effects in an interacting bosonic system have been 
an interesting subject in the context of liquid helium system for decades.
In weak disorder limit, it is believed that the disorder effects is irrelevant
to the ground state properties due to the strong repulsive interaction between
bosons in two and three dimensional systems 
\cite{huang,stringari_disorder,nelson97,lopatin02}. Similar conclusion also
applies to one-dimensional (1D) bosonic system \cite{giamarchi}.
However, when the disorder strength increases, it is argued by Fisher
{\it et al.} that the system can undergo a quantum phase transition to a 
"Bose-glass" phase \cite{fisher}, which breaks the usual $U(1)$ symmetry
of bosons to have a condensate but is insulating due to disorders.
However, the existence of the "Bose-glass" phase in the liquid helium system
is still unclear in the present experimental data \cite{reppy}.

However, the invention of optical lattice for trapping ultracold atoms
provides another route to study the disorder effects to the bosonic
systems. Experimentalists uses laser of different frequency to produce an 
artificial random potential (speckle pattern) and study the ground
state or transport properties of cold atoms \cite{zoller,optical_lattice}.
Recently, the onset of Bose glass regime is observed near the Mott-insulator
regime \cite{bose_glass}.

In this paper, we extend our earlier work \cite{early_work}
and provide a more detailed study on the origin, properties, and
effects of disorder potential in atomic waveguide system.
We show a first principle and quantitative theory to describe the disorder
potential in the atomic waveguide, and 
demonstrate that even a very small wire shape fluctuation 
can cause a strong disorder field,
which has a correlation function that vanishes at small and infinite
wavevectors and is peaked at a wavevector, $k_c$
with a length scale determined by
the atom-wire distance ($d$).  We note that similar analytical works and
numerical comparison with experimental data have been also 
studied recently \cite{aspect1,aspect2} after our work.
In strong disorder regime, the condensate becomes fragmentized
as observed in the experiment \cite{aaron1,aaron2,zimmermann1,kraft02,hinds}.
We then concentrate on 
the dynamical response of the condensate in the shaking experiment
and propose that a quantum phase 
transition from a superfluid state to an insulating
Bose glass state may be observed in the parameter regime
of current experiments. Our results 
provide a good starting point of the disorder effects to the
coherent transport and to the low dimensional Bose systems
in the future development.

This paper is organised as follows:
In Section \ref{disorder_sec} we present the theory of the
magnetic disorder potential generated from the wire shape 
distortion in the microchip
system. In Section \ref{dynamics_sec}, we study the condensate fragmentation 
in strong disorder region. Then we discuss the finite temperature
effects and other details in Sec. \ref{discussion} and summarize our results
in Section \ref{summary}.  

\section{Disordered magnetic field}
\label{disorder_sec}

In Fig. \ref{microchip}(a) we show 
a typical experimental setup of an atomic waveguide:
ultra-cold atoms are loaded into the microtrap, which 
is composed by a radiant magnetic field 
gradient (generated by the electric current
in a one-dimensional microfabricated copper wire) and 
a uniform bias field, $B_\perp$, in the transverse direction ($y$).
Atoms are confined at the potential minimum with a distance 
$d=\frac{2I_0}{cB_\perp}$ to the wire center (in c.g.s. unit), 
where $B_\perp$ exactly
cancels the field of the wire current.
A uniform offset field, $B_\|$, is applied parallel in the
wire direction ($z$) to reduce the trap loss by polarizing the 
atom spins. Additional magnetic field gradients can be also applied 
in $z$ direction to confine the atoms in the longitudinal direction.
The elongated confinement potential can be approximated by
a harmonic potential, with confinement frequencies, $\omega_\perp$ and
$\omega_\|$ in $x-y$ and $z$ directions respectively.

\subsection{General properties of disordered magnetic field}

Before studying the disorder magnetic field generated by the wire
shape fluctuations, we think it is helpful to study some more general
properties of the disorder magnetic field.
Using the fact that the typical Zeeman energy 
of atoms is very large ($\sim$ MHz),
we can safely assume that all atoms are condensed in the lowest spin
state (i.e. fully polarized in the direction of the total magnetic field) 
so that the confinement potential, $U({\bf r})$, is proportional to 
the total magnetic field:
\begin{eqnarray}
U({\bf r})=\mu_a|{\bf B}_{\rm tot}({\bf r})|,
\label{U_B}
\end{eqnarray}
where $\mu_a$ is the magnetic dipole moment of atoms and
${\bf B}_{\rm tot}({\bf r})=B_\|\hat{z}+B_\perp\hat{y}+{\bf B}_0({\bf r})
+\delta{\bf B}({\bf r})$; $\delta{\bf B}({\bf r})$ is the
disorder magnetic field, which will be calculated in details later.
Since the condensate in the microchip system is highly elongated in $z$
direction with a very small transverse radius, $R_{TF}\sim 1-5$ $\mu$m,  
we can neglect the finite size effects for simplicity and just consider
the disorder potential at the confinement center 
$(x,y,z)=(d,0,z)$, where ${\bf B}_\perp=B_\perp\hat{y}$ cancels
the unperturbed azimuthal field, ${\bf B}_0$. In the limit of small
disorder magnetic field, one can make an expansion of Eq. (\ref{U_B})
and obtain:
\begin{eqnarray}
\delta U(z)&\sim& s_\|\mu_a\delta B_z(z)+\frac{\mu_a
\left[|\delta {\bf B}(z)|^2-2\delta B_z(z)^2\right]}{2 |B_\| |},
\label{delta_U}
\end{eqnarray}
where $s_\|\pm 1$ is the sign
of the offset field, $B_\|$, in $z$ direction.
Note that in Eq. (\ref{delta_U}) the first (dominant) term
is {\it linearly} proportional to
$\delta B_z$, so that $\delta U(z)$ changes
sign if the direction of either the electric current 
or the offset magnetic field
is changed. This simple observation explains the experimental results
observed in Ref. [\onlinecite{kraft02}], where Kraft {\it et al.}
find that the positions of local potential maximum/minimum can be
exchanged by changing the direction of $B_\|$ or the current.
The second term of Eq. (\ref{delta_U}), however, does not change
sign for different $B_\|$ or current direction, and therefore 
explain why the density profile of the condensates
obtained by opposite directions of current (or opposite $B_\|$)
are not symmetric. However,
since $|B_\| |$ is in general much larger than the disorder field
($|\delta {\bf B}/B_\| |\ll 0.1\%$ in general), 
we will simply neglect the second term of Eq. (\ref{delta_U})
and consider $\delta B_z$ only when applying to the realistic disorder field
calculation, although we will still derive the disorder magnetic field
in all the components of $\delta {\bf B}$ later, which may be
useful in considering the finite size effects of BEC in the future study.

\begin{figure}

 \vbox to 7cm {\vss\hbox to 5.cm
 {\hss\
    {\includegraphics{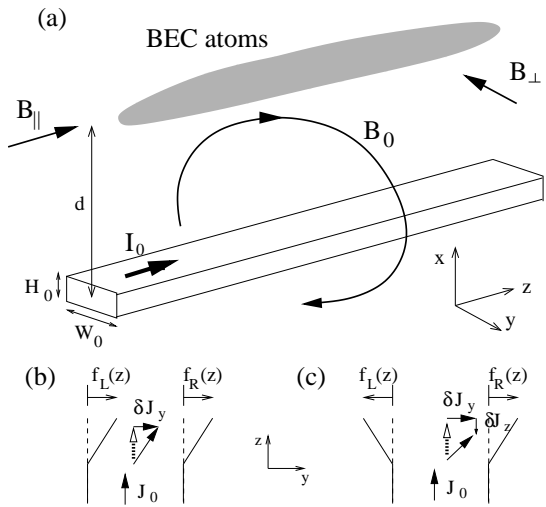}
   }
  \hss}
 }
\caption{
(a) Schematic figure of a BEC loaded in a
microtrap. ${\bf B}_0({\bf r})$, $B_\|\hat{z}$, and $B_\perp\hat{y}$
are the azimuthal magnetic field generated by a steady current,
the offset field, and the bias field respectively.
(b) and (c) present the $C-$ and $W-$ types of wire shape fluctuations
respectively.
}
\label{microchip}
\end{figure}

In this paper, we consider a microchip system where the
magnetic field is generated from a rectangular cooper wire with width
$W_0$ (in the substrate plane) and hight $H_0$ (vertical
to the substrate plane, see Fig. \ref{microchip}(a)). 
The typical scale for the semiconductor wire is $H_0\sim 1-5$ $\mu$m, and
$W_0\sim 3-50$ $\mu$m \cite{aaron1,zimmermann1}. In general, the wire 
shape fluctuation can occur in both of these two directions
due to the fluctuations during the crystal growth. However,
in this paper we assume the width fluctuation or center
changes only in the horizontal direction ($y$), but not in the vertical
direction ($x$) for simplicity (see Fig. \ref{microchip}). 
This is a good approximation because
the current density in $x$ direction should generate much smaller
magnetic field at the position right above the wire according to
the Biot-Savart law in classical electrodynamics.
Such approximation is also confirmed by the numerical work in 
Ref. [\onlinecite{aspect2}] after our first analytical study
\cite{early_work}. Therefore we can define
the wire shape function, $y=-W_0/2+f_L(z)$ for the left boundary and
$y=W_0/2+f_R(z)$ for the right boundary of the wire in the substrate ($y-z$)
plane, and the current fluctuation becomes in $y-z$ directions only, i. e.
\begin{eqnarray}
\delta {\bf J}(y,z) &\sim& \delta J_y(y,z)\hat{y}
+\delta J_z(y,z)\hat{z},
\label{delta_J_rec}
\end{eqnarray}
which are closely related to the shape of the wire and will
be studied in details below.

When studying the current fluctuation in the wire, 
it is in general tentative to approximate
the disorder nature of current fluctuation by the following expectation 
values \cite{henkel}
\begin{eqnarray}
\langle \delta {\bf J}({\bf r}) \rangle_{\rm dis} &=& 0
\label{cor_delta_J1}
\\
\langle \delta J_\alpha({\bf r})\delta J_\beta({\bf r}') \rangle_{\rm dis} 
&=& \delta_{\alpha\beta}D({\bf r}-{\bf r}'),
\label{cor_delta_J2}
\end{eqnarray}
where $\alpha,\beta=x,y,z$,
$\langle\cdots \rangle_{\rm dis}$ is disorder ensemble average,
and $D({\bf r}-{\bf r}')$ is the correlation function of the 
fluctuating current. However, we point out that
Eq. (\ref{cor_delta_J2}) is {\it not} correct for a normal conduction
wire, because it assumes no correlation between different components
of the fluctuating current. For any static current distribution,
as we will see below, the microscopic continuity equation still
applies to the disorder current so that any current fluctuation in one direction
should somehow affects the current in other directions.
Besides, the total current following in any single wire (i.e. for any specific
system as done in the realistic experiment) must be a constant too,
not {\it only} the ensemble-averaged value is unchanged as implied by
Eq. (\ref{cor_delta_J1}). Therefore the realistic current fluctuation should
obey some more strict conditions than those shown in 
Eqs. (\ref{cor_delta_J1}) and (\ref{cor_delta_J2}). This fact
will change the disorder nature of the magnetic field significantly
as we will show in the rest of this section.

\subsection{Current conservation and the boundary
conditions}

The first step is to introduce the current conservation and
the electro-dynamical equations in a self-consistent
method. We start from the static version of continuity equation, 
$\nabla\cdot\delta {\bf J}({\bf r})=0$, which gives
\begin{eqnarray}
\frac{\partial\delta J_y(y,z)}{\partial y}
+\frac{\partial\delta J_z(y,z)}{\partial z}=0.
\label{eq_J1}
\end{eqnarray}
Another equation can be derived from the Maxwell equation,
$\nabla\times{\bf E}=-\partial {\bf B}/\partial t$, with
$\partial {\bf B}/{\partial t}=0$ for static current and assuming
a constant conductivity, $\sigma$, 
inside the wire (and zero outside the wire) (i.e.
${\bf J}=\sigma{\bf E}$), so that $\nabla\times{\bf J}=0$ or 
equivalently:
\begin{eqnarray}
\frac{\partial\delta J_z(y,z)}{\partial y}-
\frac{\partial\delta J_y(y,z)}{\partial z}=0.
\label{eq_J2}
\end{eqnarray}
Eqs. (\ref{eq_J1}) and (\ref{eq_J2}) are the two equations we have to solve
by incorporating the proper boundary conditions.

The first boundary condition is from the total current
conservation in $z$ direction, which gives (See Fig. \ref{microchip}(c))
\begin{eqnarray}
I_0 &=& \int_{H_0/2}^{H_0/2} dx'\int_{-W_0/2+f_L(z)}^{-W_0/2+f_R(z)} 
dy' {\bf J}(y',z)\cdot\hat{z}
\nonumber\\
&=& H_0\int_{-W_0/2+f_L(z)}^{-W_0/2+f_R(z)} 
dy' \left(J_0+\delta J_z(y',z)\right)
\nonumber\\
&\sim& I_0 +H_0J_0\left(f_R(z)-f_L(z)\right)
\nonumber\\
&&+H_0\int_{-W_0/2}^{-W_0/2}dy'\delta J_z(y',z),
\end{eqnarray}
where $J_0\hat{z}$ is the unperturbed (or average) current density
in the wire, $I_0=J_0W_0H_0$ is the total current, and in the last
equation we have expand the right hand side by assuming small
wire shape fluctuation, i.e. 
\begin{eqnarray}
|f_{L/R}(z)|\ll W_0
\label{small_f1}
\end{eqnarray}
In other words, we have
\begin{eqnarray}
J_0\left(f_R(z)-f_L(z)\right)
&=&-\int_{-W_0/2}^{-W_0/2}dy'\delta J_z(y',z).
\label{bound_con1}
\end{eqnarray}
The second boundary condition can be obtained by requiring the current flow
is parallel to the wire boundary, whose tilted angle from the
$z$ axis is given by $\partial f_{L/W}(z)/\partial z$ for the
left and right boundary. Therefore we have (see Fig. \ref{microchip}(b))
\begin{eqnarray}
\frac{\partial f_{L/R}(z)}{\partial z}
&=& \frac{\delta J_y(\mp W_0/2+f_{L/R}(z),z)}{J_0+\delta
J_z(\mp W_0/2+ f_{L/R}(z),z)}
\nonumber\\
&\sim & \frac{\delta J_y(\mp W_0/2,z)}{J_0},
\label{bound_con2}
\end{eqnarray}
where we have used the approximation of small wire shape 
fluctuation (Eq. (\ref{small_f1})) and assumed the shape fluctuation
is smooth, i.e.
\begin{eqnarray}
|f'_{L/R}(z)|\ll 1,
\label{small_f2}
\end{eqnarray}
(Of course we always have $|\delta J_{y,z}|\ll J_0$.)
It is easy to show that Eq. (\ref{bound_con2}) can directly
imply Eq. (\ref{bound_con1}) if we take the $z$ derivative of
the later and use the continuity equation, Eq. (\ref{eq_J1}).
This shows that Eqs. (\ref{eq_J1}), (\ref{eq_J2}), and (\ref{bound_con2})
are the three independent equations we have to use to solve this problem
self-consistently. The current conservation and boundary
conditions has been included fully in the limit of small wire shape 
fluctuation (Eq. (\ref{small_f1}) and (\ref{small_f2})).

\subsection{Solution of current density fluctuations}

To solve above equations, Eqs. (\ref{eq_J1})-(\ref{eq_J2}),
we define an auxiliary function $G_J(y',z')$
which gives the fluctuating current density as follows
\begin{eqnarray}
\delta J_y(y,z)&=&J_0\frac{\partial G_J(y,z)}{\partial z}
\\
\delta J_z(y,z)&=&-J_0\frac{\partial G_J(y,z)}{\partial y}.
\label{J_yz}
\end{eqnarray}
Therefore the current continuity equation, Eq. (\ref{eq_J1}), has
been satisfy automatically, and
then $G_J(y,z)$ must satisfy a Laplacian equation,
$\nabla^2 G_J(y,z)=0$ inside the wire according to Eq. (\ref{eq_J2}).
We can use the standard method of separation of variables:
\begin{eqnarray}
G_J(y,z) &=& \frac{2}{L}\sum_{k> 0}\left(A_k e^{ky}+B_k e^{-ky}\right)
\nonumber\\
&&\ \ \ \times\left(C_k\cos(kz)+D_k\sin(kz)\right),
\label{G_J_general}
\end{eqnarray}
so that the boundary conditions (Eq. (\ref{bound_con2})) become:
\begin{eqnarray}
G_J(\pm W_0/2,z)&=&f_{R/L}(z).
\label{G_f}
\end{eqnarray}
Defining $f_{R/Z}(z)=\frac{2}{L}\sum_{k>0}\left[\tilde{f}^{R/L}_{1,k}\cos(kz)
+\tilde{f}^{R/L}_{2,k}\sin(kz)\right]$, we can solve $G_J(y,z)$ via
Eq. (\ref{G_f}) after a straightforward algebra:
\begin{widetext}
\begin{eqnarray}
G_J(y,z) 
&=& \frac{4}{L}\sum_{k>0}\left\{\frac{\cos(kz)}{\sinh(k W_0)}\left[
\sinh(kW_0/2)\cosh(ky)\tilde{f}^C_{1,k}
+\cosh(kW_0/2)\sinh(ky)\tilde{f}^W_{1,k}\right]\right.
\nonumber\\
&&+\left.\frac{\sin(kz)}{\sinh(k W_0)}\left[
\sinh(kW_0/2)\cosh(ky)f^C_{2,k}
+\cosh(kW_0/2)\sinh(ky)\tilde{f}^W_{2,k}\right]\right\},
\label{G_J_soltuion}
\end{eqnarray}
\end{widetext}
where $\tilde{f}^{C/W}_{1(2),k}\equiv\frac{1}{2}(\tilde{f}^R_{1(2),k}\pm
\tilde{f}^L_{1(2),k})$ is the cosine(sine) Fourier components 
of the wire center/width fluctuations (see Fig. \ref{microchip}(b)-(c)).
Therefore we can solve the
charge density fluctuation directly by taking the derivative
as shown in Eq. (\ref{J_yz}) in the limit of small wire
fluctuations.

\subsection{Disordered magnetic field}

Applying the approximation of small wire shape fluctuation and
above results for current fluctuations, we can obtain
the disorder vector potential, $\delta {\bf A}({\bf r})$, to be
\begin{widetext}
\begin{eqnarray}
\delta {\bf A}({\bf r}) &=& {\bf A}({\bf r})-{\bf A}_0({\bf r})
\nonumber\\
&=& \frac{1}{c}\int_{-H_0/2}^{H_0/2} dx'\int_{-\infty}^{\infty}dz'
\int_{-W_0/2+f_L(z)}^{W_0/2+f_R(z)}dy'
\frac{\delta J_y(y',z')\hat{y}
+(J_0+\delta J_z(y',z'))\hat{z}}
{\sqrt{(x-x')^2+(y-y')^2+(z-z')^2}}
-{\bf A}_0({\bf r})
\nonumber\\
&\sim&
\frac{1}{c}\int_{-H_0/2}^{H_0/2} dx'\int_{-\infty}^{\infty}dz'
\int_{-W_0/2}^{W_0/2}dy'
\frac{\delta J_y(y',z')\hat{y}+\delta J_z(y',z')\hat{z}}
{\sqrt{(x-x')^2+(y-y')^2+(z-z')^2}}
\nonumber\\
&&+\frac{1}{c}\int_{-H_0/2}^{H_0/2} dx'\int_{-\infty}^{\infty}dz'
\left[\frac{f_R(z)J_0\hat{z}}
{\sqrt{(x-x')^2+(y-W_0/2)^2+(z-z')^2}}
-\frac{f_L(z)J_0\hat{z}}
{\sqrt{(x-x')^2+(y+W_0/2)^2+(z-z')^2}}
\right]
\nonumber\\
&=&-\frac{J_0}{c}\int_{-H/2}^{H/2} dx'\int_{-\infty}^{\infty}dz'
\int_{-W_0/2}^{W_0/2}dy'
\frac{G_J(y',z')\left((z-z')\hat{y}-(y-y')\hat{z}\right)}
{\left((x-x')^2+(y-y')^2+(z-z')^2\right)^{3/2}},
\label{delta_A_rec}
\end{eqnarray}
where we have used Eq. (\ref{G_f}) to cancel the extra terms via 
integration by parts.
The magnetic field can then be obtained directly: 
\begin{eqnarray}
\delta{\bf B}({\bf r})
&=&\frac{J_0}{c}\int_{-\infty}^{\infty}dz'\int_{-H_0/2}^{H_0/2} dx'
\int_{-W_0/2}^{W_0/2}dy'G_J(y',z')
\left\{-\frac{2(x-x')^2-(y-y')^2-(z-z')^2}
{\left((x-x')^2+(y-y')^2+(z-z')^2\right)^{5/2}}\hat{x}\right.
\nonumber\\
&&\left.+\frac{-3(x-x')(y-y')}
{\left((x-x')^2+(y-y')^2+(z-z')^2\right)^{5/2}}\hat{y}
+\frac{-3(x-x')(z-z')}
{\left((x-x')^2+(y-y')^2+(z-z')^2\right)^{5/2}}\hat{z}\right\}
\label{delta_B_rec}
\end{eqnarray}
\end{widetext}

Note that Eqs. (\ref{delta_U}), (\ref{G_J_soltuion}) and (\ref{delta_B_rec}) 
are the main results of this section, using the following approximation: 
(1) the wire boundary fluctuation in the vertical ($x$)
direction can be neglected, and (2) the wire shape fluctuations
in $y$ directions are small, i.e. 
$|f_{L/R}(z)|\ll W_0$ and $|f'_{L/R}(z)|\ll 1$. We believe these 
are reasonable 
approximations in the regime of present experimental interest.

\subsection{Finite wire size effect and disorder correlation function}
\label{finite_size}

Although one can numerically calculate the disorder field from
Eq. (\ref{delta_B_rec}) as done in Ref. [\onlinecite{aspect1,aspect2}], 
but we may further approximate it by using
the fact that (i) the wire hight, $H_0$, is usually much smaller than $d$
in the $x$ direction, and (ii) in practice we just need to study $\delta B_z$
at $(x,y,z)=(d,0,z)$ according to the earlier discussion about Eq. 
(\ref{delta_U}). We then evaluate $\delta B_z(d,0,z)$ by integrating
the wire width and obtain
\begin{widetext}
\begin{eqnarray}
\delta B_z(d,0,z) &=& \frac{H_0J_0}{c}\int_{-\infty}^{+\infty}dz'
\int_{-W_0/2}^{W_0/2}dy'\frac{-3G_J(y',z')d(z-z')}{(d^2+y'{}^2+(z-z')^2)^{5/2}}
\nonumber\\
&=&\frac{-6H_0J_0}{cd}\int_{-\infty}^{+\infty}dz'\frac{(z-z')}
{(1+(z-z')^2)^{5/2}}
\sum_{n=0}^\infty\frac{(-1)^n5\cdot 7\cdot \cdot(2n+3)}
{n!\cdot 2^n (1+(z-z')^2)^n}
\int_0^{W_0/2d}G_J(y'd,z'd)y'{}^{2n}dy',
\end{eqnarray}
where we have changed to a dimensionless dummy variable inside the
integration. The last integral can be evaluated directly to be
(the odd part of $y$ in $G_J(y,z)$ has been integrated out to be zero)
\begin{eqnarray}
\int_0^{W_0/2d}G_J(yd,zd)y^{2n}dy
&=&\frac{4}{L}\sum_{k>0}\frac{\sinh(kW_0/2)}{\sinh(kW_0)}
\left[\cos(kdz)f^C_{1,k}+\sin(kdz)f^C_{2,k}\right]
\frac{\gamma(2n+1,kW_0/2)-\gamma(2n+1,-kW_0/2)}{2(kd)^{2n+1}},
\nonumber\\
\end{eqnarray}
\end{widetext}
where $\gamma(n,x)\equiv\int_0^x x^{n-1}e^{-x}dx$ is the incomplete Gamma
function. Therefore we can calculate the disorder correlation function,
$\langle\delta U(z)\delta U(z')\rangle_{\rm dis}$, whose Fourier component in
momentum space, $\Delta_k=\int dz\,e^{-ik(z-z')}
\langle\delta U(z)\delta U(z')\rangle_{\rm dis}$, is
\begin{eqnarray}
\Delta_k&=&\left(\frac{2I_0\mu_a}{c}\right)^2
\frac{(kd)^4}{d^4}|D(kW_0,kd)K_1(kd)|^2 F_k,
\label{Delta_k}
\end{eqnarray}
where
\begin{eqnarray}
D(x,y)&\equiv &\frac{2\sinh(x/2)}{x\sinh(x)}
\sum_{n=0}^\infty \frac{(-1)^n}{n!\cdot (2y)^n}\frac{K_{n+1}(y)}{K_1(y)}
\nonumber\\
&& \times\left[\gamma(2n+1,x/2)-\gamma(2n+1,-x/2)\right],
\end{eqnarray}
and $F_k\equiv\int dz\,e^{-ik(z-z')}\langle f_C(z)f_C(z')\rangle$.
$K_i(x)$ is the modified Bessel function of the second kind.

\begin{figure}

 \vbox to 8.7cm {\vss\hbox to 5.cm
 {\hss\
   {\includegraphics{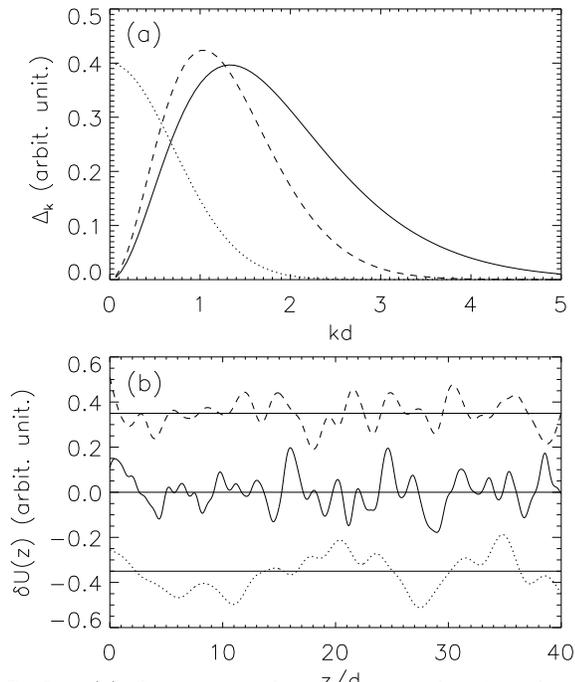}
   }
  \hss}
 }
\caption{
(a) Calculated disorder correlation function, $\Delta_k$. 
Solid and dashed lines are for $F_k$=constant
and $F_k\propto e^{-k^2\eta^2}$ with $\eta=d/2$ respectively. 
For comparison, we also show
a Gaussian correlation function (dotted line) of length scale $d$.
(b) Typical disorder potentials in real space, calculated 
from the disorder correlation function shown in (a). 
Note that the scales of the vertical axes in both figures are
not the same for different curves, up to the overall strengths of disorders.
}
\label{u_x0}
\end{figure}

\subsection{Disorder strength}

To measure the strength of disorder, we define the strength $u_s$ to be
$u_s^2 \equiv \langle u(z)u(z)\rangle=\int\frac{dk}{2\pi}\Delta_k$, 
which is obtained by integrating the whole spectrum of 
the correlation function, $\Delta_k$.
We then obtain
\begin{eqnarray}
u_s^2 &=&\left(\frac{2I_0\mu_a}{c}\right)^2\frac{1}{\pi d^4}
\int_0^\infty dk (kd)^4K_1(kd)^2
D(kW_0,kd)^2F_k
\nonumber\\
&=&\left(\frac{2I_0\mu_a}{cd}\right)^2\frac{1}{\pi d^3}
\int_0^\infty dp p^4K_1(p)^2
D\left(\frac{pW_0}{d},p\right)^2F_{p/d}
\nonumber\\
&\equiv&
\left(\frac{2I_0\mu_a}{cd}\right)^2\frac{S(W_0,d;\{F_k\})}{\pi d^3},
\label{u_s}
\end{eqnarray}
where $p=kd$ is dimensionless variable, and
a new function, $S(W_0,d;\{F_k\})$, is defined to
incorporate the wire shape disorder, which is
proportional to $F_0$ in large $d$ limit. Therefore, if considering $W_0\ll d$
limit, we may simplify above result further to be
\begin{eqnarray}
u_s &=& C\cdot\frac{2I_0\mu_a}{cd}\cdot\left(\frac{F_{0}}{d^3}\right)^{1/2}
=C\mu_a B_\perp\cdot\left(\frac{F_{0}}{d^3}\right)^{1/2}
\label{u_s2}
\end{eqnarray}
where $C$ is of order of one and
depends on $W_0$ and disorder length scale only weakly for large $d$.
According to Eq. (\ref{u_s2}), for a fixed current ($I_0$), the
disorder strength $u_s$ scales as $d^{-2.5}$
for large $d$, roughly consistent with the experimental result
shown in Ref.\cite{kraft02}, where
they estimated the disorder field
$\sim d^{-2.2}$ by equating the chemical potential to the disorder field
at the onset of the fragmentation.
Besides, Eq. (\ref{u_s}) shows that even if the wire center fluctuation
is very small ($F_0\ll d^3$), 
it can still generates a large disorder potential compared to
the BEC chemical potential, $\mu$, because
the energy of the bias magnetic field, $\mu_a B_\perp$, is 
in general of order of MHz, much larger than $\mu$, which is
just of order of kHz. Therefore it seems almost impossible to
reduce the disorder effect in the microchip experiment 
by improving the wire sample, and one has to consider the disorder
effect to the coherent transport more seriously especially
when $d$ is not large.

\subsection{Model wire fluctuation and numerical results}

\begin{figure}

 \vbox to 8.7cm {\vss\hbox to 5.cm
 {\hss\
    {\includegraphics{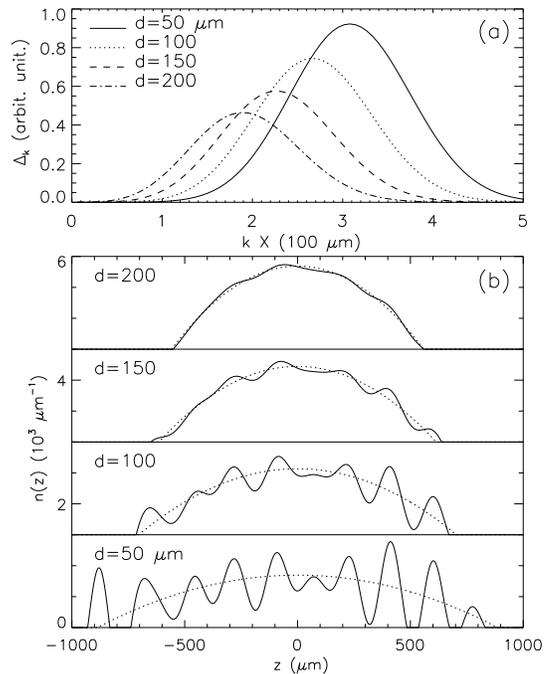}
   }
  \hss}
 }
\caption{
(a) Calculated disorder correlation function, $\Delta_k$, for
different atom-wire distance $d$.
(b) Calculated density profiles
of a $^{23}$Na condensate by
using the same parameters as the experiments (Fig. 4 of Ref. \cite{aaron1}).
Dotted lines are the density without disorder potential.
In both figures, $F_k$ is approximated by the form described 
in the text. 
}
\label{u_x}
\end{figure}
In Fig. \ref{u_x0}(a) we show the numerical results of $\Delta_k$
with $W_0=0$ for simplicity (the results are 
similar for $W_0<0.1 d$). 
Solid(dashed) line is for 
$F_k=$constant($F_k\propto e^{-k^2\eta^2}$ with $\eta/d=0.5$).
Results of a disorder potential with a Gaussian correlation function 
(i.e. $\Delta_k\propto e^{-k^2d^2}$, which can{\it not} be a result
of a current conserving calculation as we discussed earlier)
is also shown in the same figure (dotted line) for comparison.
Unlike the usual Gaussian correlation most adapted in the
literature,  $\Delta_k$, obtained in 
Eq. (\ref{Delta_k}), is zero at $k=0$ and peaked at a finite wavevector
$k_0\sim 1.33/d$ (solid line) even if the wire center fluctuation has 
{\it zero} correlation length (i.e. $F_k$=constant). 
In other words such length scale can be completely {\it generic} and is 
mainly determined by the microtrap geometry
rather than by the length scale of the underlying surface disorder
of the wire.
This interesting results are due to the current conservation inside the
wire, which requires the current fluctuation to be cancelled 
in low wavelength limit ($kd\ll 1$). The finite width of the peak
shows that such disorder can be considered as a Gaussian correlated
random potential. 
In Fig. \ref{u_x0}(b), we show typical disorder potentials in real 
space based on the disorder correlation functions discussed above.
One can see that the usual Gaussian-type correlation (dotted line)
can{\it not} give any periodicity in the disorder potential.
On the other hand, if we assume a Gaussian correlation for
the wire center fluctuation, i.e. $F_k\propto e^{-k^2\eta^2}$ (dashed lines),
it will help to reduce the contribution from wavevectors higher than
$1/d$, and makes the quasi-periodicity more transparent as
shown in Fig. \ref{u_x0}(b). The peak position of $\Delta_k$ is then
shifted to lower value $k_1$, which gives the length scale, 
$\lambda=2\pi/k_1$, of the Gaussian correlated random potential in real space.

Another interesting situation is that if the wire shape has 
not only one random disorder but also an intrinsic 
periodic disorder with a period 
$\eta_0$ (i.e. $F_k\sim C_1 \delta_{k,k_0}+ C_2 e^{-k^2\eta^2}$,
where $k_0=2\pi/\eta_0$ and $C_{1/2}$ are relative 
strengths of these two disorders), 
Eq. (\ref{Delta_k}) then has a ``double-peak'' structure in
$\Delta_k$ at $k=k_1$ and $k_2$. The coexistence of 
these two length scales ($\eta_0$, which is $d$-independent and 
$\lambda$, which is $d$-dependent) in a BEC fragmentation 
can explain the double periodicity observed in Ref. 
[\onlinecite{zimmermann1}]. 
When temperature is raised above $T_c$, the chemical
potential may be smaller than one (stronger) disorder 
but still larger than the other (weaker) one,
so that only one length is revealed in the fragmentation of 
a thermal cloud \cite{zimmermann1}.

Although the exact form of the wire shape fluctuation
($f_C(z)$) has to be determined by a microscope as have been done
in Refs. [\onlinecite{aspect1,aspect2}] and is sample-dependent, we think it
can be generally described by assuming $F_k=\frac{\Delta_C}{2}
(e^{-(k-k_1)^2\eta_1^2}+e^{-(k+k_1)^2\eta_1^2})$, where
$\langle f_C(z)f_C(z)\rangle=\frac{\Delta_C}{2\eta_1\sqrt{\pi}}$
measures the wire shape fluctuations. For $k_1=0$, we have a normal
(Gaussian) distribution function of $f_C(z)$ with a length scale $\eta_1$.
For $k_1\eta_1>1$, $f_C(z)$ has a Gaussian correlated structure 
with a length scale,
$\lambda=2\pi/k_1$. For $\eta_1=0$, we recover the zero length scale
fluctuation (white noise) for $f_C(z)$ and then $k_1=k_0\sim 1.33/d$
as shown in Fig. \ref{u_x0}(a). In this paper, we choose our 
parameters to model the fragmentation observed in MIT group \cite{aaron1}
and will set $\langle f_C(z)f_C(z)\rangle$=(0.1 $\mu$m)$^2$, $\eta_1=
100$ $\mu$m, and $2\pi/k_1=200$ $\mu$m for the numerical calculation
in the rest of this paper.

In Fig. \ref{u_x}(a) we show the numerical results of $\Delta_k$
for different $d$ by using above model fluctuation of the wire shape.
As expected, the underlying disorder (now has a length scale
in the Gaussian correlated form) will change
the position of the peak of $\Delta_k$, while the overall 
length scale still depends on $d$.
In Fig. \ref{u_x}(b) we calculate a static BEC density profile
of sodium atoms in different disorder potential 
strengths by solving the 1D static
Gross-Pitaevskii equation (GPE):
\begin{eqnarray}
0=-\frac{\partial_z^2\Phi}{2 m}+\left[V_0
+\delta U+g_{\rm 1D}|\Phi |^2
-\mu\right]\Phi
\label{GPE}
\end{eqnarray}
within Thomas-Fermi (TF) approximation \cite{bec_book}, where 
$m$ is the atom mass, and $V_0(z)=\frac{1}{2}m\omega_\|^2 z^2$ 
is the axial harmonic confinement potential.
\begin{eqnarray}
g_{\rm 1D}&=&\frac{4 a}{m a_\perp^2},\ \ \ 
\left(a_\perp=\sqrt{\frac{2}{m\omega_\perp}}\right)
\end{eqnarray}
is the effective 1D interaction strength incorporating the radial 
confinement potential \cite{tonks,note_1D}: $a$ is the 3D 
$s$-wave scattering length
$\omega_\perp$ is the single particle confinement potential 
in the transverse mode
and $a_\perp$ is the typical radius in the transverse direction. 
We can see that the condensate becomes fragmentized when $d$
becomes small, as observed in the experiments.

\section{Fragmentation in strong disorder limit}
\label{dynamics_sec}

In this section, we focus on the strong disorder regime, where 
the condensate is fragmentized into several pieces with weak
coupling strength between the neighboring fragments
We will use the modified tight-binding model to
study the dynamical properties of 
these condensate fragments in the superfluid phase regime
by using the dynamical meanfield (Gross-Pittaivskii) theory.

\subsection{Tight-binding approximation}

For the fragments observed in the atomic waveguide experiments 
\cite{aaron1,aaron2,zimmermann1,kraft02}, it seems quiet reasonable to 
assume that the single particle tunneling between two neighboring fragments
is so weak that the system can be described by a tight-binding approximation.
However, unlike the usual tight-binding model used in the optical lattice,
the atom-atom interaction effects is very crucial in the fragment case
in determining the onsite particle wavefunction.  
We start from the full second quantization representation of 1D boson problem:
\begin{eqnarray}
i\frac{\partial\hat{\psi}}{\partial t}&=&
-\frac{\nabla^2\hat{\psi}}{2m}+\left(V_0
+\delta U-\mu\right)\hat{\psi}+g_{\rm 1D}
\hat{\psi}^\dagger\hat{\psi}\hat{\psi},
\label{qt_eq}
\end{eqnarray}
where $\hat{\psi}(z,t)$ and $\hat{\psi}(z,t)^\dagger$ are bosonic operators.
Now we introduce the onsite single particle wavefunction to describe
the discrete nature and expand the single boson operator as follows
\begin{eqnarray}
\hat{\psi}(z,t)&=&\sum_{j}\phi_{j}(z)\hat{a}_{j}(t),
\label{psi_expand}
\end{eqnarray}
where $\phi_{j}(z)$ is the onsite groundstate wavefunction of
the local potential well $j$, which is centered at $z=z_j$.
$\hat{a}_{j}$ is the bosonic operator for that specific eigenstate. 
In Eq. (\ref{psi_expand}) we have assumed that all the boson atoms in each
well are in the local groundstate, and no higher energy modes can be
excited (single band approximation) in each well.
We will discuss the validity of such approximation later.
Substituting above equation into Eq. (\ref{qt_eq}), and integrating
the $z$ coordinate after multiplying $\phi_j^\ast(z)$ in both sides of the
Eq. (\ref{qt_eq}), we obtain (after neglecting the next nearest
hopping energy and interaction between nearest neighboring wells)
\begin{eqnarray}
i\frac{\partial\hat{a}_j}{\partial t}&=&
-\frac{1}{2}\sum_{\alpha=\pm 1}K_{j,j+\alpha}\hat{a}_{j+\alpha}
+U_j\hat{a}^\dagger_j\hat{a}_j\hat{a}_j
\nonumber\\
&&+\int dz\left[-\frac{1}{2m}\phi_j^\ast\nabla^2\phi_j+\phi_j^\ast
(V_0+\delta U-\mu)\phi_j\right]\hat{a}_j,
\nonumber\\
\label{qt_eq2}
\end{eqnarray}
where $K_{j,j'}\equiv-2\left[\frac{1}{2m}\int dz\partial_z
\phi_j(z)\partial_z\phi_{j'}(z)+
\int dz\phi_j(z)\right.$ $\left.\phi_{j'}(z)(V_0(z)
+\delta U(z)-\mu)\right]$ is tunneling 
amplitude between $j$th to $j'$th well. 
$U_j\equiv g_{\rm 1D}\int dz |\phi_j|^4$
is the onsite charging energy.

Note that if we use
the {\it noninteracting} ground state, $\phi_{j0}$, for each well 
to expand the single particle operator in Eq. (\ref{psi_expand}),
we obtain the
well-known single-band Bose Hubbard model (BHM) \cite{zoller}:
\begin{eqnarray}
i\frac{\partial\hat{a}_j}{\partial t}&=&
-\frac{1}{2}\sum_{\alpha=\pm 1}K_{j,j+\alpha}\hat{a}_{j+\alpha}
+(V_{0,j}+\delta U_j-\mu)\hat{a}_j
\nonumber\\
&&+U_j\hat{a}^\dagger_j\hat{a}_j\hat{a}_j,
\label{BHM}
\end{eqnarray}
where $V_{0,j}\equiv\int dz V_0|\phi_{j0}|^2$ and 
$\delta U_{j}\equiv\int dz \delta U|\phi_{j0}|^2$ is the onsite
smooth potential and disorder potential. (The local kinetic energy,
i.e. the first term in the second line of Eq. (\ref{BHM}),
can be absorbed into the chemical potential.) 
The approximation of such {\it noninteracting wavefunction-based} 
single band BHM is self-consistent only if the onsite
excitation energy, $\omega_{loc}$, is much larger than any other 
energy scale in Eq. (\ref{BHM}), $K_{j,j+1}$, $\delta U_j$, $V_{0,j}$ $U_j$, 
{\it and} $N_j U_j$, where $N_j=\langle a^\dagger_j a^{}_j\rangle$ 
is the average
onsite number of atoms. The last criterion ($\omega_{loc}\gg N_jU_j$)
is required to confirm that the atom-atom interaction does 
not change the onsite wavefunction from the noninteracting one, $\phi_{j0}$,
which determines the values of $K_{j,j+1}$, $U_j$, and $\delta U_j$ etc.
as shown above. If $\omega_{loc}$ is of the same order as $N_jU_j$,
the interaction between atoms in each well can deform their wavefunction
from the noninteracting ground state, $\psi_{j0}$, by including higher
energy onsite eigenstates. As a result, the single band approximation
used in Eq. (\ref{psi_expand}) fails.

However, in our present condensate fragment system, the number of particles
in a fragment is so large ($N_j\sim 10^{4-5}$) that the criteria, $\omega_{loc}
\gg N_jU_j$ cannot be satisfied in most of the experimental situations
(for the fragment size, we can calculate that typically $U_j\sim 0.05$ Hz
and $\omega_{loc}\sim 10-50$ Hz). As a result,
it is still reasonable to assume that atoms in each fragment can be still
described by a single (condensate) wavefunction, 
$\phi_j$, which, different from the
noninteracting wavefunction $\phi_{j0}$, is determined mainly by 
the competition between the atom-atom interaction (instead of the 
local kinetic energy) and local confinement provided by 
the disorder potential. Therefore we can assume the $U(1)$
symmetry is broken in each well, so that 
the onsite wavefunction, $\phi_j$, can be determined
by the following local static GP equation:
\begin{eqnarray}
0=-\frac{1}{2m}\nabla^2\phi_j+
(V_0+\delta U-\mu)\phi_j+g_{\rm 1D}{N}_j^0|\phi_j|^2\phi_j,
\label{onsite_GPE}
\end{eqnarray}
where $\phi_j$ is normalized to one and 
${N}_j^0$ is of the equilibrium number of particles in
well $j$. Using above equation to eliminate the 
last term of Eq. (\ref{qt_eq2}),
we obtain
\begin{eqnarray}
i\frac{\partial\hat{a}_j}{\partial t}&=&
-\frac{1}{2}\sum_{\alpha=\pm 1}K_{j,j+\alpha}\hat{a}_{j+\alpha}
+U_j\hat{a}^\dagger_j\hat{a}_j\hat{a}_j-{N}_j^0\hat{a}_j^\dagger\hat{a}_j.
\nonumber\\
\label{qt_eq2_2}
\end{eqnarray}

Note Eq. (\ref{qt_eq2_2}) is different from Eq. (\ref{BHM}), because the
onsite wavefunction now have to be determined by including interaction, 
local kinetic energy, and local trapping potential via 
Eq. (\ref{onsite_GPE}). As a result,
the criterion of single band approximation used in the regular
BHM (Eq. (\ref{BHM})) can be softened to be 
$\omega_{loc}\gg K_{j,j+1},U_j$ for Eq. (\ref{qt_eq2_2}).
At the same time we still assume the temperature is low enough that
the whole system is in quasi-condensate region, i.e. local density 
fluctuation in each well and the phase difference between 
neighboring wells can be neglected in determining the equilibrium onsite
wavefunction. 
The randomness of the disorder potential is now absorbed into the
the randomness of $N_j^0$ and also $\phi_j$, which makes $U_j$
and $K_{j,j'}$ are also random. We note that a simplified 
Hamiltonian of strong disorder
in both diagonal and off-diagonal term (i.e. without interaction $U_j$ of 
Eq. (\ref{qt_eq2_2})) has been also studied recently in 
Ref. [\onlinecite{real_RG}].

The dynamics of Eq. (\ref{qt_eq2_2}) can be studied within its
meanfield version by taking $\langle\hat{a}_j(t)\rangle
=\langle\hat{a}_j^\dagger(t)\rangle=N_j(t)e^{iS_j(t)}$,
where $N_j$ and $S_j$ are the number of particle and
phase function of the $j$th condensate fragment respectively:
\begin{eqnarray}
\dot{N}_j&=&-\sum_{\alpha=\pm 1}K_{j,j+\alpha}\sqrt{N_jN_{j+\alpha}}
\sin(S_{j+\alpha}-S_j)
\label{eqn_N}
\\
\dot{S}_j&=&\frac{1}{2}\sum_{\alpha=\pm 1}K_{j,j+\alpha}
\sqrt{\frac{N_{j+\alpha}}
{N_j}}\cos(S_{j+\alpha}-S_j)
\nonumber\\
&&-(N_j-{N}_j^0)U_j
\nonumber\\
&\approx&-(N_j-{N}_j^0)U_j,
\label{eqn_S}
\end{eqnarray}
where we have neglected the single particle tunneling term for 
simplicity, which
can be shown to be much smaller than the onsite charging energy (see below).
It is interesting to note that Eqs. (\ref{eqn_N})-(\ref{eqn_S}) 
are very similar
to the well-known equations for Josephson junction \cite{JJ}. 
However, the interaction between
bosonic atoms leads to an effective onsite charging energy (the last term of
Eq. (\ref{eqn_S})), which closes the equations by changing the 
number of particles
and phase functions simultaneously.
Note that throughout this paper we will neglect the dynamics of 
the quantum fluctuation
completely for simplicity, because it is very small due to the large average 
number of particles in each fragment ($N_j^0> 10^4$). 

\subsection{Probability distribution of 
parameters in the tight-binding model}
\label{distribution}

\begin{figure}

 \vbox to 6.5cm {\vss\hbox to 5.cm
 {\hss\
   {\includegraphics{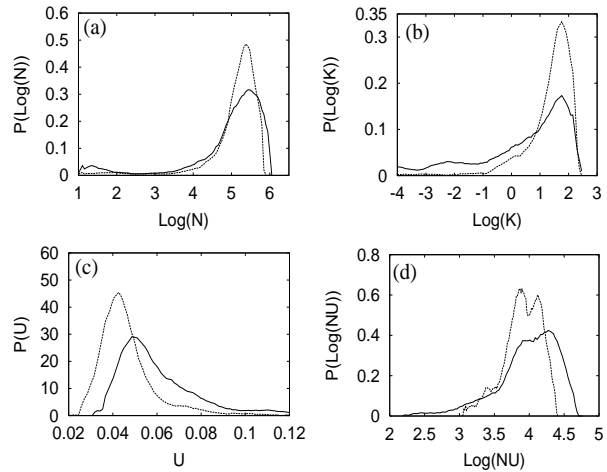}
   }
  \hss}
 }
\caption{(a)-(d) are respectively the
probability distribution of local number of atoms $N_j^0$, tunneling energy, 
$K_{j,j+1}$, onsite charging energy, $U_j$, and onsite effective disorder
potential $N_j^0U_j$ (see Eq. (\ref{qt_eq2}). Dotted (solid) lines are
for disorder strength $s=u_s/u_s^0=0.5 (1.2)$ respectively. The units of 
energy variables are normalized to Hz ($\hbar=1$).
}
\label{P_parameter}
\end{figure}

In Eqs. (\ref{eqn_N})-(\ref{eqn_S}) the parameters, $K_{j,j+1}$, $U_j$ and 
${N}_j^0$, are all random numbers, because the disorder potential in the 
microchip may generate fragments of different sizes and at different (random)
positions, which affects the
tunneling amplitude and onsite charging energy as well. 
In our calculation, we use a Gaussian trial wavefunction to solve the
onsite wavefunction, $\phi_j(z)$, of Eq. (\ref{onsite_GPE}) at 
each potential $j$,
keeping the chemical potential fixed by the {\it total} 
number of particles of the
{\it whole} condensate. The single particle tunneling amplitude $K_{j,j+1}$, 
onsite interaction energy, $U_j$, and number of atoms in $j$th fragment
$N_j^0$, are calculated by $\phi_j(z)$ as in the standard method \cite{zoller}.
In principle these three quantities are {\it not} independent of each other,
but it is still useful and instructive to study their probability 
distribution function individually for a given disordered magnetic field
studied in the previous section. 
In Fig. \ref{P_parameter}, we show the distribution
function of them for relatively weak ($s=0.5$ dotted line) and strong 
($s=1.2$ solid line)
disorder potential, where $s\equiv u_s/u_s^0$ and $u_s^0$ is the
disorder strength defined in Eq. (\ref{u_s}) with the
same disorder realization as used in Fig. \ref{u_x} at $d=100$ $\mu$m.
This is calculated by considering $2\times 10^6$ number of sodium atoms
in ten potential wells for each time and then averaging the results 
over different (more than twenty thousands) disorder realizations.

Several features can be observed in Fig. \ref{P_parameter}:
(i) when disorder is enhanced, the distribution functions of all parameters
become broadened as expected and also change their mean values.
(ii) For the number of particles per site 
(Fig. \ref{P_parameter}(a)), stronger potential wells can trap more atoms,
leaving only very small number ($< 10^2$ per site) 
of atoms in the weaker potential wells. Therefore in the fragment
situation we are considering (Eqs. (\ref{onsite_GPE})-(\ref{qt_eq2})),
some weaker potential wells can have only very few atoms, which behave
like a weak link between their two neighboring wells.
This can be observed in Fig. \ref{P_parameter}(b) that (iii) when the
disorder strength increases, more ``weak'' junctions between fragments
appear so that the whole system may becomes more closer to an
insulating phase. Simultaneously, we find that (iv) the onsite charing energy
is also increased by disorder as shown in Fig. \ref{P_parameter}(c).
This is due to the fact that stronger onsite disorder potential 
can reduce the size
of the fragment, leading to stronger onsite interaction energy.
(v) Finally, in Fig. \ref{P_parameter}(d), we show the distribution function of
the ``effective onsite disorder potential'', $N_j^0U_j$, as derived in
Eq. (\ref{qt_eq2_2}). The double peak structure in weak disorder limit
(dotted line) is due to the different contributions of $N_j^0$ and $U_j$ 
respectively. When disorder increases, they merge to one peak and move 
to stronger side. Therefore we can summarize that the disorder potential 
of the microchip fragmentation can increase the inhomogeneity of 
the density profile, reduce the tunneling amplitude, $K_{j,j+1}$,
increase the onsite charging energy, $U_j$, and also increase onsite potential
variation. All of these features lead to a strong indication of an 
insulating phase in strong disorder region. In the rest of this section,
We will discuss
the dynamics of such system under the shaking experiment and
study how it can be related to the superfluid to insulator transition.

\subsection{Two-fragment dynamics and self-trapping}

We start with the two-fragment case ($j=1,2$), where
some unique dynamical properties due to the nonlinear GPE can be described
more precisely. 
For the two fragment case, the only eigen frequency is
$\omega_J = \sqrt{2\tilde{K}\bar{U}}$ ($\tilde{K}\equiv 
K_{1,2}\sqrt{N_1^0N_2^0}$ and $\bar{U}\equiv\frac{1}{2}(U_1+U_2)$)
for the small oscillation of
$\Delta N\equiv N_2-N_1-(N_2^0-N_1^0)$ and $\Delta S\equiv S_2-S_1$.
It is easy to show that Eqs. (\ref{eqn_N})-(\ref{eqn_S}) are
equivalent to the problem of single planar pendulum, where $\Delta S(t)$ and
$\Delta N(t)$ represent the angular position of the weight from the vertical 
line and the angular velocity respectively (we assume 
$\sqrt{N_1N_2}\sim\sqrt{N_1^0 N_2^0}$ for small density variation).
\begin{figure}

 \vbox to 9.3cm {\vss\hbox to 5.cm
 {\hss\
   {\includegraphics{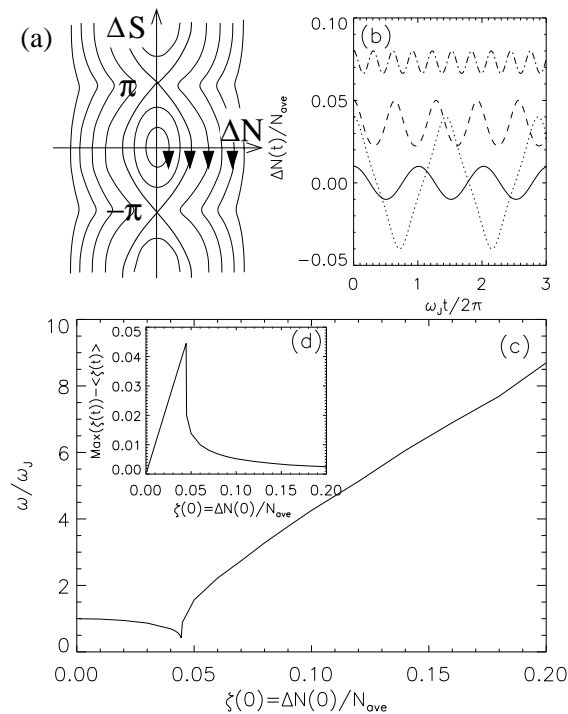}
   }
  \hss}
 }
\caption{Dynamics of two identical fragment system:
(a) Flow diagram in the phase space. (b) Time evolution of 
the density variation with different initial $\Delta N$. 
(c) Oscillation frequency and (d) amplitude (about its average value) 
as a function of initial density modulation from equilibrium value.  
}
\label{wJ_DN2}
\end{figure}
In such simple two-fragment situation, 
an interesting phenomenon, ``self-trapping'' effect
can be observed in the two-fragment case when $\Delta S$ is
larger than $\pi$. As shown in the phase portrait of
Fig. \ref{wJ_DN2}(a), the system trajectories of the initial conditions
denoted by the first two
right arrows do {\it not} pass the origin ($\Delta N=0$) and keep
flow away by growing $\Delta S$ exponentially. 
In Fig. \ref{wJ_DN2}(b), we show a numerical results for $\Delta N(t)$
from different initial $\Delta N(0)$ (with $\Delta S(0)=0$), using 
a set of typical parameters: 
$N_1^0=N_2^0=10^5$, $K_{1,2}=1$ Hz and $\bar{U}=0.05$ Hz. 
We can see that when initial displacement, $\Delta N(0)/N_1$, is less
than 1\% (solid line), the number of atoms oscillates between these
two fragments with frequency $\omega_J=100$ Hz. When $\Delta N(0)$
increases, the oscillation amplitude first increases accordingly 
(dotted line) and then suddenly decreases for $\Delta N(0)> \Delta N_c
\sim 0.04 N_1$ (dashed and dash-dotted lines). 
Besides, the density imbalance $\Delta N(t)$ then never passes zero but
just oscillates about a new average value.
Such counterintuitive result is due to the nonlinear nature of
Eq. (\ref{eqn_N}), and can be realized 
in a single pendulum problem where the weight has a 
very large initial velocity to overcome the 
gravitation potential even at the highest position of 
the circle ($\Delta S=\pi$) 
and then keep rotating forward 
with a nonzero velocity (see the trajectory following the 
first two right arrow in Fig. \ref{wJ_DN2}(a)). 
It is also very similar to the Ac Josephson effect, where a constant
electric potential drop between the two connecting superconductors
can cause a oscillating current between them. In our present 
situation, the initial potential drop is provided by the large imbalance
of number of particles so that the fast oscillation 
of single particle tunneling
cannot reduce such initial imbalance except for other damping mechanism.
We can calculate the critical amplitude of the initial density variation
easily and obtain
\begin{eqnarray}
|\Delta N_c |=2\sqrt{\frac{2\tilde{K}}{\bar{U}}}.
\end{eqnarray}
We note that such critical density variation becomes smaller when 
the tunneling energy is weaker and/or the average onsite 
charging energy is larger, i.e. close to the insulating phase. 

In Fig. \ref{wJ_DN2}(c)-(d), we plot the oscillation frequency and 
amplitude as a function of different initial density variation, 
$\zeta(0)=\Delta N(0)/N_{\rm ave}$. It is easy to see that the
oscillation amplitude drops very fast when initial
displacement is larger than the self-trapping point.
This result also exists even 
when considering the full quantum mechanics
in such simple two well system \cite{anatoli}, because
the typical number of particles per site in the fragments is so 
large ($>10^4$) that the quantum fluctuation can be neglected.

\subsection{Shaking experiment in condensate fragments}

In optical lattice system, the dynamics of a condensate cloud can be studied
by the ``shaking experiment'', a suddenly shifting 
of the global confinement potential with a finite 
displacement, and then observing the successive center of mass motion
\cite{bishop,shaking_exp}. When the initial displace is small, the
condensate oscillates harmonically, showing a coherent Josephson junction
tunneling between neighboring wells. When the displacement is larger 
than some critical value, however, the center of mass motion becomes 
strongly damped, indicating a dynamical instability, which is a classical
phase transition due to the breakdown of meanfield solution \cite{bishop}.
It is believed that when the strength of optical lattice 
is tuned to be strong enough, the shaking experiment with small 
displacement can be used to investigate the proposed superfluid to (Mott)
insulator transition \cite{zoller,SF-MI}, 
where (unlike in the superfluid phase) 
the small displacement should not result
in a coherent center of mass motion \cite{anatoli_dw}. 
In our earlier work \cite{early_work}, 
we proposed that similar experiments can be done  
to investigate the superfluid to insulator transition in the microchip
fragments, where the insulating phase is 
best understood as a Bose glass phase due to the underlying 
disorder nature \cite{bose_glass}. Here we will study such multi-well 
dynamics in more details via the meanfield equation of motion derived 
in Eqs. (\ref{eqn_N})-(\ref{eqn_S}).

\begin{figure}
 \vbox to 9.5cm {\vss\hbox to 5.cm
 {\hss\
   {\includegraphics{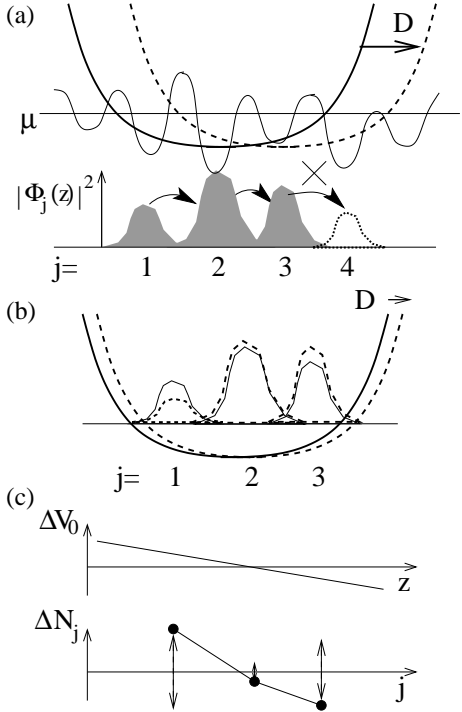}
   }
  \hss}
 }
\caption{Schematic potential and density profile of condensate fragments
in a microchip system. (a) Self-trapping effects for large displacement,
$D$. The atoms in fragment 3 cannot tunneling as a condensate 
into the potential
well 4, which is generated by the large displacement of global confinement
potential. The noisy curve
in the upper portion denotes the local disorder potential, and the lower 
portion shows the fragment density profile before shaking.
Solid/dashed lines denote the global confinement 
potential before/after shaking.
(b) Density variation for small displacement (no disorder potential is 
shown here).The dashed line for the density profile denotes the new 
equilibrium profile in the new potential after shaking.
Solid lines are the density profile before shaking.
(c) The potential and density variation with respect to the new
potential and equilibrium density profile. Such $\Delta N_j$ becomes
the source of fragment dynamics as time evolves.
Up-down arrows indicate the possible density oscillation in each well.
}
\label{edge_effect}
\end{figure}

However, although there are many similarity between the condensate fragments
we consider here and the condensates in optical lattice 
\cite{zoller,optical_lattice,bose_glass}, their
difference in the sizes of local potential wells does bring
some significant difference of their dynamics. We first note that
due to the large size of the disorder potential well in the microchip,
the overall condensate density profile is far away from the result
of Thomas-Fermi inverse paraba (compared the solid and dotted lines in 
Fig. \ref{u_x}(b)). Actually, one can expect that the density profile
has a kind of ``discontinuity'' around the edge of the whole condensate.
Applying the self-trapping dynamics discussed in previous section, we
then expect that the atoms in the edge fragments will 
{\it not} effective tunneling
into the potential well next to it, which is created by 
shifting the confinement potential (see Fig. \ref{edge_effect}(a)).
Such strong edge effects is right due to the nonlinearity-induced
self-trapping phenomena as discussed earlier. Therefore we do not expect 
that the condensate fragments will be driven to motion as a whole
by such shaking experiment. The only possible exception is by thermalizing
the atoms in the edge fragments to higher energy modes, which is certainly
not a coherent motion at all and is not our current interest in this paper.

Despite of such huge difference, the shaking experiment 
with small displacement can be still applied to
study the coherent motion in the condensate fragments by realizing
the density profiles of the whole condensate. 
As schematically shown in Fig. \ref{edge_effect}(b),
although the small displacement $D$ may not change the whole condensate
position with respect to the disorder potential, it is indeed 
capable to change the onsite potential strength to a slightly new
value, and hence the atoms in each fragments become to flow between
neighboring wells to response the change of local chemical potential.
Precisely speaking, such small displacement shift should change the
shape of local potential well also, and hence change their position as
well as well width/depth. However, we believe it is a reasonable
approximation to neglect
the change of well positions and widths, and concentrate on 
the effects mainly from the deviation of local chemical potential, which
gives a new equilibrium density profile, $N_j^{new}$, so that
the old density profile, $N_j^{old}$, becomes an initial non-equilibrium
source for the coming dynamics. In Fig. \ref{edge_effect}(c), we schematically
depict the linear change of (parabolic) confinement potential and hence the
unbalanced density source for the future dynamics, $\Delta N_j\equiv
N_j^{old}-N_j^{new}$. If the whole condensate is in superfluid regime,
we expect that $\Delta N_j(t)\equiv N_j(t)-N_j^{new}$ will oscillate
about zero with a definitely phase relation with respect to its
neighboring site (vertical dashed arrows in Fig. \ref{edge_effect}(c)),
when $|\Delta N_j(t)|$ is small enough.
In general, the density variation 
($\zeta_j(t)\equiv |N_j(t)-N_j^0|/N_j^0$, where $N_j^0\equiv N_j^{new}$ and
$N_j(t=0)=N_j^{old}$)
is largest at the edge fragments, because the potential deviation is
the largest there while its average number of atoms ($N_j^0$) is in general 
smaller than the fragments in the center of condensate.

In the following calculation of fragment dynamics, we shift the constant
change of chemical potential so that $\Delta V_0(z)=
\frac{m}{2}\omega_\|^2(z+D)^2-\frac{m}{2}\omega_\|^2z^2=m\omega_\|^2Dz
\equiv {\cal D}z$
is zero at the center of the whole condensate with an effective potential 
displacement ${\cal D}$. The initial density variation (i.e. the density 
{\it before} shaking, $N_j^{old}$, respect to the new equilibrium
density, $N_j^{new}$, {\it after} shaking) can be approximated
by $\Delta N_j(t=0)\sim \Delta V_0(z_j)/U_j$. We will therefore consider
how the density variation ($\Delta N_j(t)$) evolves as a function of time
with different displacement, ${\cal D}$. 
Besides, we also make a further approximation by assuming the global
confinement potential can be neglected when calculating the onsite wavefunction
and other dynamical variables, because they are mainly determined by
the local disorder potential.

\subsection{Numerical results}

\begin{figure}

 \vbox to 9cm {\vss\hbox to 5.cm
 {\hss\
   {\includegraphics{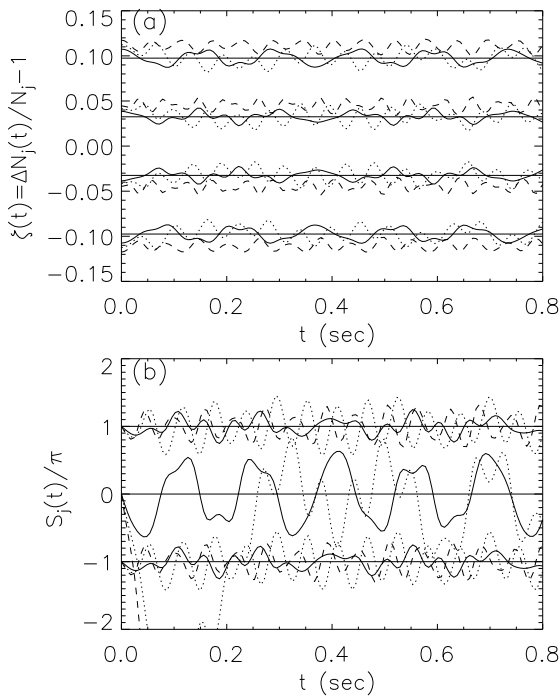}
   }
  \hss}
 }
\caption{
(a) Density and (b) phase oscillation in each fragments of a typical
four fragment condensate.
In both figures, solid, dotted, and dashed lines are results for
$\cal D$=0.15, 0.2, and 0.3 Hz/$\mu$m respectively. In (a)
the four sets of curves are for well $j=1$, 2, 3, and 4 respectively
from bottom to top. In (b) the three sets of curves are phase gradients
in the neighboring sites: $S_2-S_1$,
$S_3-S_2$, and $S_4-S_3$ respectively from bottom to top.
}
\label{NS_t2}
\end{figure}

As an example of the condensate fragment dynamics, we show in Fig.
\ref{NS_t2} the typical results for a four fragment system. We choose
$K_{1,2}=K_{3,4}=1$ Hz $K_{2,3}=0.5$ Hz, $U_{1,2,3}=0.05$ Hz, and
$N_1^0=N_4^0=1.3\times 10^5$ and $N_2^0=N_3^0=1.2\times 10^5$ as 
some typical values shown in Section \ref{distribution}. 
Several features can be observed from Fig. \ref{NS_t2}. (i) When the
displacement is small (solid lines), both density profiles and phase
gradients oscillate about their new equilibrium values with small amplitude. 
(ii) When the displacement is increased, one of the phase gradient (here it 
is $S_3-S_2$) becomes unbound, leading to a self-trapping phenomena
where the density profile oscillates around a nonzero mean value 
(dashed lines).
(iii) Between the generic oscillation in small displacement and
the self-trapping in the large displacement, we find that the 
condensate dynamics becomes chaotic in the intermediate range of 
displacement (dotted lines). Such chaotic dynamics can be investigated 
via the frequency spectrum (see Ref. [\onlinecite{early_work}]) 
or the time correlation function.
This is a classical instability of the meanfield GPE, and its 
result to the center of mass motion in optical lattice has been
investigated in Refs. [\onlinecite{bishop}]. 

\begin{figure}

 \vbox to 5cm {\vss\hbox to 5.cm
 {\hss\
   {\includegraphics{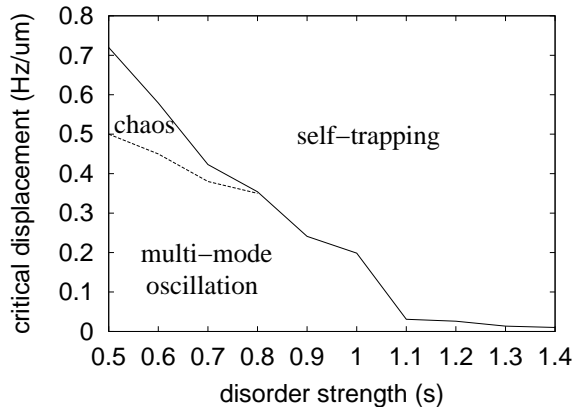}
   }
  \hss}
 }
\caption{Critical shaking displacement of a ten-fragment system as a function 
of disorder strength ($s=u_s/u_s^0$). Total number of atoms is $2\times 10^6$.
}
\label{D_c}
\end{figure}

In Fig. \ref{D_c} we plot a numerical calculated dynamical phase diagram
for ten-fragment system with $N_{tot}=2\times 10^6$ sodium atoms in the
disorder potential calculated earlier (at $d=100$ $\mu$m) 
in terms of disorder strength ($s$) and critical 
displacement, $D$. Here the dashed line 
separates the multi-mode oscillation (e.g. solid lines in Fig. \ref{NS_t2}) 
from the chaotic motion (e.g. dotted lines in Fig. \ref{NS_t2}), while
the solid line 
separates the multi-mode/chaotic motion from the self-trapping motion
(e.g. dashed lines in Fig. \ref{NS_t2}). We can see 
that the critical displacement of self-trapping
decreases dramatically for $s=u_s/u_s^0<1.1$ and then becomes smoothly
for $s>1.1$. It never reaches zero in our meanfield approximation.
We believe that after including the full quantum fluctuation,
the meanfield solution becomes incorrect in the strong disorder region,
where the quantum fluctuation can decrease the critical displacement
so much that even infinitesmall displacement to the 
condensate will be self-trapped
without coherent motion (i.e. the solid line of Fig. \ref{D_c}
terminates at finite disorder strength, say $s=1.1$) \cite{anatoli_dw}.
This can be understand as a signature of quantum phase transition from
superfluid phase in small disorder to an insulating phase in 
strong disorder, which is best understood as a Bose glass phase \cite{fisher}
due to the nature of randomness. However, since the low energy excitation
properties in such strong disorder regime are still poorly understood, 
we could not exclude the possibility of different quantum phases,
such as Mott glass as suggested by Giamarchi {\it et. al.} 
\cite{giamarchi_mott}
due to the competition between a (white noise) random potential
and a commensurate periodic potential.

\begin{figure}

 \vbox to 6.8cm {\vss\hbox to 5.cm
 {\hss\
   {\includegraphics{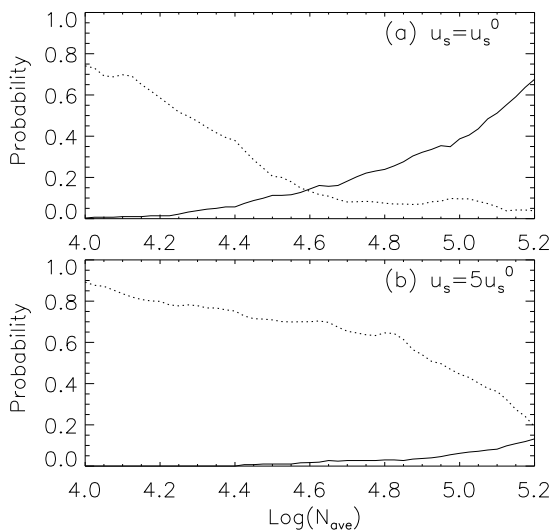}
   }
  \hss}
 }
\caption{
The dotted line shows the probability {\cal P}($Q < 1$). 
The system is in the superfluid state
when this probability is close to zero (see text),
while it is in an insulating state when 
 it is close to one. The solid line shows the probability that one can observe
fluctuations in the number of atoms between the neighboring wells
with fluctuations in the number of particles larger than
$\zeta_{\rm min}=0.1$ and oscillation frequency 
larger than $\omega_{\rm min}=2\pi\times 1$ Hz. When this
probabilities approaches zero, the system will appear self-trapped in the
shaking experiments. Here $N_{\rm ave}$ is the average number of atoms
in a single well (mini-condensate). Disorder strength for (a) are
the same as used in Fig. \ref{u_x}b for $d=100$ $\mu$m (denoted to be
$u_s^0$), while it is five times stronger in (b).
The results are averaged over more than twenty thousand pairs of two-fragment
system with average number of atoms, $N_{\rm ave}$, per well.
}
\label{P_zeta}
\end{figure}

\subsection{Estimate quantum fluctuation effects}

The transition from superfluid phase to insulator phase is driven by 
the quantum fluctuation in the competition between interaction and
random potential \cite{fisher}. Including the quantum fluctuations
in the dynamics of condensates has been studied in some limited cases
\cite{qt_dynamics,qt_dynamics2} by either solving the full coupled
nonlinear Gross-Pitaevskii-Bogoliubov-de-Gennie equation \cite{qt_dynamics},
using dynamical meanfield variational method \cite{zoller}
or by using truncated Wigner approximation \cite{qt_dynamics2}.
Studies including random potential is much more difficult and
still in progress. Here we will give some estimate 
about such transition using the statistical properties,
which, at least in principle, should be able to be probed by the
dynamical shaking experiment when displacement is tuned to infinitesmall.

Transition into the insulating state may be characterized by the
ratio of the ``Josephson energy'' between the neighboring wells to the
charging energy of one
of the wells, i.e. 
$Q_j\equiv\frac{8{K}_{j,j+1}N_j^0N_{j+1}^0}{(U_j+U_{j+1})/2}$. 
Without disorder potential, the meanfield calculation
\cite{ehud,sachdev} estimates the superfluid (SF) to Mott insulator (MI)
transition at $Q=1$.
For the Gaussian correlated random potential we discuss here,
we expect that quantum fluctuation induced 
SF to BG phase transition \cite{note2} should also appear at 
roughly the same place. Due to the randomness nature of the condensate 
fragments, the effects of quantum fluctuation is important
if the probability to 
have $Q_j<1$ ($\equiv{\cal P}(Q<1)$),
is of the order of one, where the probability if obtained by averaging
over disorder ensemble.
In Fig. \ref{P_zeta} we show the calculated ${\cal P}(Q<1)$
as a function of the atom density for
two different strengths of the disorder potential (dotted lines). 
In order to capture the quantum fluctuation nature of each 
random junction, we calculate $Q$
for each pair of fragments across the potential barrier
with average number of atoms $N_{\rm ave}$ per well.
According to Fig. \ref{P_zeta}, we can estimate that the quantum 
fluctuations are crucially important to a pair of fragments
when $N_{\rm ave}< 10^{4.4}\sim 2.5\times 10^4$ for $s=u_s/u_s^0=1$.
This critical average number of particles become higher
($N_{\rm ave}< 10^{5}$) 
for stronger disorder potential ($s=u_s/u_s^0=5$).
In the same figure, we also show the calculated probability 
(solid lines) to have 
Josephson frequencies $\omega_J$ to be larger than some
minimum frequency, $\omega_{\rm min}=2\pi\times 1$ Hz, {\it and}
the density contrast, $\zeta=\frac{\Delta N}{N_0}$, to be larger
than $\zeta_{\rm min}=10\%$.
The values of $\omega_{\rm min}$ and $\zeta_{\rm min}$ are 
given by the experimental resolution of frequency and density
deviations \cite{private} for the density oscillation.
We can see that this probability (denoted by 
${\cal P}(\omega>\omega_{\rm min},\zeta>\zeta_{\rm min})$) 
decreases when average number of atoms per
well decreases and/or the disorder strength increases (toward the insulating
phase). Both of these two results (${\cal P}(Q<1)$ and
${\cal P}(\omega>\omega_{\rm min},\zeta>\zeta_{\rm min})$) suggest
that one can observe the superfluid to insulator (Bose glass) transition
\cite{note2} in the parameter regime of present experiments.

It is interesting to compare above estimate of quantum fluctuations
(Fig. \ref{P_zeta})
with the dynamical phase diagram associate with the shaking experiments
shown in Fig. \ref{D_c}. In Fig. \ref{D_c}, the dynamical properties 
are calculated for an ensemble of ten fragments, instead of two.
Since the superfluid dynamics of the whole condensate 
can be strongly suppressed if one of these junctions (say, junction
$l$ between fragment $l$ and $l+1$)
becomes insulating due to strong quantum fluctuation ($Q_l<1$),
we can roughly estimate that the criterion of superfluid to insulator
transition occurs when $M_{\rm junc}{\cal P}(Q<1)$ is of order of one,
where $M_{\rm junc}$ is number of total junctions. From the data
shown in Fig. \ref{P_zeta}(a), we can see that this value is about
$9\times 0.05=0.45$, for $N_{\rm ave}\sim \log(2\times 10^5)=5.3$.
Therefore we find that the quantum fluctuations of a ten-fragment
condensate with $N_{\rm ave}\sim 2\times 10^5$ per well should become
strong when the disorder strength is about (or slightly larger than) 
$u_s^0$. This is consistent with the estimate (see previous subsection)
via the self-trapping phenomena observed in a meanfield dynamics 
associate with shaking experiments shown in Fig. \ref{D_c}.

Finally we note that the results shown in Fig. \ref{P_zeta}(b) are
obtained by the same effective 1D interaction strength used in
Fig. \ref{P_zeta}(a).
In the realistic experiment, however, $g_{\rm 1D}$ will be changed
if one tunes the wire current and the bias field, $B_\perp$ 
simultaneously (in order to keep condensate at the same position).
This is because the effective radial confinement energy will be also 
increased during such process. However, it is easy to show that
one can reduce the confinement energy effectively
by increasing the off-set magnetic field, $B_\|$, simultaneously
without changing any other system parameters.
Therefore the final system can be kept almost the same as
the one before tuning, except the disorder field has been increased or
decreased independently via the composed process mentioned above.

\section{Finite temperature effects}
\label{discussion}

To estimate the validity of our calculation in finite temperature regime,
we take the following four steps: 
(i) first we have to clarify the general 
concept about the temperature effects in low dimensional BEC system.
(We do not need to include any disorder potential at this point.)
Due to the finite size effects in the longitudinal direction, it is possible
that a boson system becomes 1D quantum degenerate (or say the lowest energy
eigenstate becomes macroscopically occupied, see Ref. [\onlinecite{ketterle}]) 
when the temperature is below the degeneracy
temperature, $T_d= N\omega_\|/\ln(2N)$, where $N$ is the total number
of atoms and $\omega_\|$ is the confinement frequency in the longitudinal
direction. When considering the interaction-induced thermal phase fluctuation
effects for temperature below $T_d$, Petrov {\it et. al.} \cite{petrov}
show that a true condensate, where both density and phase fluctuations
are small within the finite system size, can be achieved when 
temperature is below another temperature scale, 
$T_{ph}=T_d\times\omega_\|/\mu$, where $\mu$ is the chemical potential.
For a temperature between these two temperatures, i.e.
$T_{ph}<T<T_d$, the system has a frozen density fluctuation but
finite thermal phase fluctuation. The typical experimental
parameters (as we consider in this paper in a superfluid regime),
we can estimate that $T_d\sim 13.2$ $\mu$K, and $T_{ph}\sim 127$
nK (for $N=10^6$, $\omega_\|=2\pi\times 4$ Hz, and that chemical potential
is about 3 kHz for atom-wire separation being about 100 $\mu$m).
The present experiments were done at a temperature of about 100 nK
\cite{private}. Therefore, we can safely say
that the condensate we consider here (similar to the present 
experimental conditions of MIT group) is in the true condensate regime
(or deep in the quasi-condensate regime), where thermal fluctuation 
is very small or the thermal fluctuation length is comparable
to the whole system size.

(ii) Secondly, we consider the presence of a strong disorder potential,
where we approximate the fragmented condensate by a single band 
Bose-Hubbard model and solve its dynamics in meanfield approximation.
The first approximation can be easily justified in our system from
the following two reasons: first, we can turn
on the disorder potential (by moving the atom cloud closer to the 
wire) adiabatically, so that the ground state remains
in condensate without high-energy excitations. Actually, the 
effective temperature compared to the band gap becomes even smaller
due to the shrink of band energy, since no thermal reservoir is connected
to the atom cloud trapped in the magnetic field (thermal fluctuation
from the wire surface \cite{henkel} 
can be neglected here because of the larger atom-wire
separation). This is very similar to the situation in an optical lattice.
Secondly, each fragment itself is in the 
true condensate regime due to large number of atoms ($\sim 10^{4-5}$)
and stronger confinement provided by the correlated disorder 
potential. The critical temperature estimated for a single fragment
is about $T_{ph}\sim 1$ $\mu$K, well above the temperature quoted in
the experiments. Therefore, the thermal 
excitation to a higher energy band is strongly suppressed and can be
safely neglected. 

(iii) However, the second approximation, the meanfield approximation
for the dynamical motion of condensates, is self-justified only when 
in the deep superfluid regime and when the temperature is lower than 
both the Josephson energy $\tilde{K}_{j,j'}$ and the local chemical
potential deviation, $\delta\mu=U_j(N_j-N_j^0)$ (see 
Eqs. (\ref{eqn_N})-(\ref{eqn_S})), which 
are the only two energy parameters in a meanfield version of the Hubbard
model. Using the typical parameters 
we tracked from the data of the MIT group,
the above criteria can be safely fulfilled due to the large number of
atoms per fragment.

(iv) Finally, we apply such a single band Hubbard model to study
the quantum phase transition by increasing the disorder potential strength
(Fig. \ref{D_c}) or by reducing the number of atoms per well 
(Fig. \ref{P_zeta}).
When considering the quantum fluctuation effects near the transition
point, we note that the temperature has to be below the charging 
energy, $U_j$, instead of $N_jU_j$ in the classical limit, in order
to see the quantum effects. Such a condition, however, may not be 
satisfied in our system. In other words, the finite temperature
effects may smoothen the transition/crossover from the superfluid
phase to the insulating Bose glass phase, but such crossover should still 
be controlled by the zero temperature quantum critical point.
Therefore, we believe our results, like Fig. \ref{P_zeta} and dynamical
properties near the insulating phase, should still be
qualitatively valid at finite temperature.

\section{Summary}
\label{summary}

In summary, we study in detail the disorder magnetic potential in an atomic
waveguide (microchip) and quantitatively 
explain the fragmentation phenomena observed
in the experiments. 
Our results shows that the disorder effects can be very strong even for a 
very small wire edge fluctuation, and hence not
negligible in most of the present experimental situations.
We also study the dynamical properties
of an array of fragments (self-trapping, multi-pole oscillation and 
modulational instability), and propose a superfluid-to-insulator
crossover in strong disorder limit, which can be probed by a shaking experiment
within the experimentally accessible regime. 

\section{Acknowledgement}

The authors thank useful discussion with E. Demler, A.E. Leanhardt, S. Kraft,
L.D. Lukin, D.E. Pritchard, G. Rafel, E. Altman, and W. Hofstetter.
We especially acknowledge the critical discussion with A.E. Leanhardt
for the experimental details.



\begin{thebibliography}{99}
\bibitem{note2}
Note that in the thermodynamic limit,
various regimes may be distinguished only for a bounded disorder,
while in a finite system there is no
true phase transition --- however, one may still observe
the sharp crossovers.
\end{thebibliography}
\end{document}